\documentclass[aps,amssymb,amsmath,twocolumn, pra, superscriptaddress]{revtex4}
\usepackage{graphicx}
\usepackage{epsfig}
\usepackage{dcolumn}% Align table columns on decimal point
\usepackage{bm}% bold math
\usepackage{bbm}
\usepackage{ulem}
\usepackage[colorlinks, citecolor=blue, urlcolor=blue]{hyperref}
\usepackage[up]{subfigure}
\usepackage{color}
\usepackage{soul}

\usepackage{braket}
\usepackage{comment}

\newcommand{\be}{\begin{equation}}
\newcommand{\ee}{\end{equation}}
\newcommand{\bc}{\begin{center}}
\newcommand{\ec}{\end{center}}
\newcommand{\bi}{\begin{itemize}}
\newcommand{\ei}{\end{itemize}}
\newcommand{\bea}{\begin{eqnarray}}
\newcommand{\eea}{\end{eqnarray}}

\usepackage{caption}

\begin{document}
\title{Discrete-time quantum walk algorithm for ranking nodes on a network}

\author{Prateek Chawla}
\email{prateekc@imsc.res.in}
\affiliation{The Institute of Mathematical Sciences, C.I.T. Campus, Taramani, Chennai 600113, India}
\affiliation{Homi Bhabha National Institute, Training School Complex, Anushakti Nagar, Mumbai 400094, India}

\author{Roopesh Mangal}
%\email{roopeshmangal1333@gmail.com}
\affiliation{The Institute of Mathematical Sciences, C.I.T. Campus, Taramani, Chennai 600113, India}

\author{C. M. Chandrashekar}
\email{chandru@imsc.res.in}
\affiliation{The Institute of Mathematical Sciences, C.I.T. Campus, Taramani, Chennai 600113, India}
\affiliation{Homi Bhabha National Institute, Training School Complex, Anushakti Nagar, Mumbai 400094, India}
%==================================================
\begin{abstract}

We present a quantum algorithm for ranking the nodes on a network in their order of importance. The algorithm is based on a directed discrete-time quantum walk, and works on all directed networks. This algorithm can theoretically be applied to the entire  internet, and thus can function as a quantum PageRank algorithm. Our analysis shows that the hierarchy of quantum rank matches well with the hierarchy of classical rank for directed tree network and for non-trivial cyclic networks, the hierarchy of quantum ranks do not exactly match to the hierarchy of the classical rank. This highlights the role of quantum interference and fluctuations in networks and the importance of using quantum algorithms to rank nodes in quantum networks. Another application this algorithm can envision is to model the dynamics on networks mimicking the chemical complexes and rank active centers in order of reactivities. Since discrete-time quantum walks are implementable on current quantum processing systems, this algorithm will also be of practical relevance in analysis of quantum architecture.

\end{abstract}
%====================================================
\maketitle
%\preprint{Version}

%=================================================
\section{Introduction}
%=================================================

A quantum walk is by and large a quantum mechanical analogue of a classical random walk\,\cite{Riaz58,Feynman86,ADZ93,Mayer96} without having the randomness associated with the dynamics. Quantum walks has served as a base for development of various quantum algorithms and in modeling the dynamics of many quantum systems. They are also becoming increasingly relevant  topic of interest beyond the conventional, quantum information and physics community. 
We have two well studied version of quantum walk, the continuous-time quantum walk (CTQW) and the discrete-time quantum walk (DTQW). The dynamics of the CTQW is defined only on a position Hilbert space, whereas, an additional coin Hilbert space along with the position Hilbert space is used to define the dynamics of DTQW. Both the variants have been shown to be effective at performing various quantum computational tasks\,\cite{Kempe03, NV00, IKS05, DW08, Childs09, VA12, Lovett10, CGW13}. Beyond computational tasks, CTQW has played an important role in modeling the energy transfer in photosynthetic material\,\cite{MRLA08}. The additional coin degree of freedom in DTQW has served as an extra degree of freedom to control the dynamics and model the physical phenomena, such as topological phases\,\cite{COB15, SRFL08, KRBD10, AO13}, Dirac equation and its associated dynamics\,\cite{MC16, BB94, CMC13, DP14, Perez16, CBR10, Strauch06, KBLC18}. Therefore, in terms of utility, a quantum walk is a very powerful scheme for quantum simulations, in direct analogy to the role of classical random walk in classical simulations over the past few decades. Experimental implementation of quantum walks in variety of quantum systems, such as NMR\,\cite{RLBL05}, integrated photonics\,\cite{SCP10,BFL10,Peruzzo10}, ion traps\,\cite{SRS09,ZKG10}, and cold atoms\,\cite{ KFCSWMW09} makes it a promising protocol of future quantum technologies. 
 
Complex networks have become a part and parcel of modern life and scientific research. As a consequence, there has been significant research in the field of network analysis, which is applied to not only the World Wide Web\,\cite{GGS10, CV07}, but also to social and biological  systems\,\cite{RSMOB02, BO04}. Problems pertaining to communication, storage and transport of information have seen significant interest from the scientific community, and has been studied in the form of network analysis. With the recent advances in the field of quantum information and computation, quantum networks are envisioned to dominate the architecture of all aspects of quantum information, communication and computation protocols. Some early versions of these networks have already been created and analysed\,\cite{CE04, PPM08, TQN11, LMESM10, LL09, Kimble08, Wiersma10, BDCZ98, DBCZ99, SSRG11, LMRAG11, Simon10, Lauri10}. Some of the physical models exhibit interesting properties such as long-distance entanglement\,\cite{VMDC04, PVMDC05, KY10, HDERNHB05}.

A fundamental problem in a vast network of information therefore becomes one of classification, search and retrieval. In this respect, for a complex network, ranking nodes of the network on the basis of relevance of information required becomes a challenge of some importance. A significant development in this field has been the introduction of the PageRank algorithm\, \cite{BP98, BMPW98a, BMPW98b, LM04, HK03, AM12}, which is the heart of Google's search engine. An important step in this direction for quantum networks has been the successful attempt\, \cite{PMD12} to quantize the classical protocol, based on Szegedy's scheme\,\cite{Szegedy04} for quantization of Markov chains.

It has been proven that the quantum methods for ranking nodes outperform their classical counterparts \cite{PMCMD13, SBDGGZ12, LTRMW15, HM09} on different kinds of networks, but the quantum protocol tested is based on Szegedy's scheme. Szegedy's scheme is a variant of a DTQW that does not require a coin operator, but needs an additional Hilbert space of the same dimension as its position space. Thus, the additional resource requirement makes its physical realization an uphill task compared to a standard single particle DTQW implementation where the internal degree of freedom of the coin acts as a coin Hilbert space. 

To overcome the difficulty of using an additional position space in implementing the existing quantum rank algorithms we propose a new algorithm based on a directed DTQW (D-DTQW) which will require only a position and a coin Hilbert space to implement the algorithm effectively. This algorithm will be relevant beyond ranking node in networks, to rank active centers on chemical compounds in order of reactivities, to model dynamics in complex quantum networks, and analyze the connectivity of quantum communication networks and other quantum architectures.

In order to rank the nodes of a quantum network, it becomes essential to preserve some of the quantum properties of the network in an objective, network-independent manner. In the classical protocol, this was visualized by a browser performing a random walk on the Web, which is what made Google's algorithm a success. Going by the same analogy for a quantum network, our quantum algorithm identifies the nodes of a network as states in a multi-dimensional Hilbert space, and just like its classical counterpart, performs a D-DTQW on the network. Since a single-particle DTQW has been experimentally implemented in various quantum systems, its directed variant will also be directly implementable on a physical quantum circuit, which makes the algorithm practically scalable for a network as well.

This paper is organized as follows, in section\,\ref{sec:dtqw}, we present an overview and brief analysis of our D-DTQW algorithm and how it ranks the nodes on a graph. In section\,\ref{sec:res} we present the results of our simulations and comparisons with the classical PageRank algorithm on some networks. We collate all our results and conclude with remarks in section\,\ref{sec:conc}.

%=====================================================
\section{Directed Discrete-time Quantum Walk algorithm for Ranking Nodes}
\label{sec:dtqw}
%==================================================
\subsection{Discrete-time quantum walk}
%================================================
The DTQW for a single particle walker on a one-dimensional lattice is defined as an evolution in the Hilbert space $\mathcal{H} = \mathcal{H}_C \otimes \mathcal{H}_P$, where $\mathcal{H}_C$ and $\mathcal{H}_P$ are coin and position Hilbert spaces, respectively. The coin space is taken to have the basis states $\left\{ \ket{\uparrow}, \ket{\downarrow} \right\}$ which represents the internal states of the walker. The position space is defined by the basis states $\ket{x}$, where $x \in \mathbb{Z}$. The initial state of the walker is a combination of the coin and position space state of the form, 
\be	
	\label{eq:eq2.1}
	\ket{\psi}_0  = \left( \alpha \ket{\uparrow} + \beta \ket{\downarrow}\right) \otimes \ket{x=0} \quad;\quad  |\alpha|^2 + |\beta|^2 = 1.
\ee
Here $\alpha$ and $\beta$ are the amplitudes of the coin states. The evolution operator is defined by the action of a coin operator on the coin space, followed by the action of a coin state dependent position shift operator on the entire system. The coin operator may be a U(2) matrix, but is mostly used in the single-parameter form,
\be
	\label{eq:eq2.2}
	C_\theta = \begin{bmatrix}
	~~~\cos(\theta) & -i\sin(\theta) \\
	-i\sin(\theta) & ~~~\cos(\theta)
	\end{bmatrix} \otimes \sum_x \ket{x} \bra{x}. 
\ee
The shift operator shifts different components of the probability amplitude in different directions, and is given by
\be
	\label{eq:eq2.3}
	S_x = \sum_{x \in \mathbb{Z}} \left[ \ket{\uparrow}\bra{\uparrow} \otimes \ket{x-1} \bra{x} + \ket{\downarrow}\bra{\downarrow} \otimes \ket{x+1}\bra{x} \right].
\ee
In general, the state of the walker after $n$ steps of evolution is given as
\be
	\label{eq:eq2.4}
	\ket{\psi}_n = \left( S_x C_\theta \right)^n \ket{\psi}_0.
\ee

%=================================================================
\subsection{Directed discrete-time quantum walk algorithm}
%====================================================================
In the D-DTQW\,\cite{HM09}, the shift operator allows shifting in only one direction, and can be written as the $S_+$ or $S_-$, depending on the direction in which evolution is directed towards. Thus, the D-DTQW shift operator is given by,
\be 
	\label{eq:eq2.5}
	S_\pm = \begin{cases}
				\sum_x \ket{\uparrow}\bra{\uparrow} \otimes \ket{x\pm 1}\bra{x} + \ket{\downarrow}\bra{\downarrow} \otimes \ket{x}\bra{x} & or, \\
				 \sum_x \ket{\uparrow}\bra{\uparrow} \otimes \ket{x}\bra{x} + \ket{\downarrow}\bra{\downarrow} \otimes \ket{x\pm 1}\bra{x}.
			\end{cases}
\ee
To construct an algorithm for ranking the nodes on a digraph, the D-DTQW operators used for defining the standard directed evolution need to be modified. The coin and position shift operators must be made dependent on the properties of the graph or network. Therefore, the redefined node dependent coin operation takes the from, 
\be 
	\label{eq:eq2.6}
	C = \sum_x \begin{bmatrix}
	\sqrt{\frac{1}{\alpha_x+1}} & \sqrt{\frac{\alpha_x}{\alpha_x+1}} \\
	\sqrt{\frac{\alpha_x}{\alpha_x+1}} & -\sqrt{\frac{1}{\alpha_x+1}}
	\end{bmatrix} \otimes \ket{x}\bra{x}.
\ee
Here $\alpha_x$ represents the proportion of the incoming weight compared to the total incoming and outgoing weights at the node represented by $\ket{x}$, i.e.,  $\alpha_x = \frac{d_i}{d_i+d_o}$, where $d_i$ is the indegree and $d_o$ is the outdegree of node $\ket{x}$. It is trivial to verify that $C$ is unitary. The shift operator takes the form, 
\be 
	\label{eq:eq2.7}
	S = \sum_x \left[ \ket{\uparrow}\bra{\uparrow} \otimes \ket{x}\bra{x} + \sum_k \left( \ket{\downarrow}\bra{\downarrow} \otimes U_{kx} \ket{k}\bra{x} \right) \right].
\ee 
Here the matrix $U$ is unitary, so that $S$ is also unitary. The algorithm is encoded directly into the operators as follows:\\
The coin operator is defined at each node, and essentially rotates the state depending on how much proportion of the data throughput at the particular node is incoming data. The incoming proportion is then 'stored' in the $\ket{\uparrow}$ coin state, and the outgoing information is sent to all nodes $\ket{k}$ to which the node $\ket{x}$ is connected, in a proportion that is determined by the matrix $U$, which will be defined below. 

For a directed graph, in general, the adjacency matrix $A$ is not symmetric. Therefore, we transform $A$ into a scattering matrix by considering the singular value decomposition of $A$ as,
\be
	\label{eq:eq2.8}
	A = P\Lambda Q,
\ee
where $P$ and $Q$ are the matrices of the left and right singular eigenvectors of $AA^{*}$ and $A^{*}A$, respectively. The matrix $\Lambda$ is a diagonal matrix consisting of the eigenvalues, which essentially contains information about how much information goes through each node, and in this sense, acts like a transfer matrix. The $\Lambda$ is Hermitian, and is converted into a unitary form as $S = e^{i\Lambda}$. Since $A$ was a square matrix, $P,Q$ and $S$ are all square matrices of the order of $A$. In addition, it may be verified that all the three matrices $P,Q$ and $S$ are unitary by themselves, and thus, the operator
\be
	\label{eq:eq2.9}
	U = P e^{i\Lambda} Q,
\ee
will also be unitary. Physically, this makes the walk behave as if the $\ket{\downarrow}$ part of the probability is being scattered off each node and being redistributed among the directed edges.

So far, we have not mentioned anything about the weights of the edges connecting the nodes. In a more general case when the edges are weighted, the adjacency matrix $A$ takes care of the weights on its own, and so the shift operator is left unchanged. The coin operator is changed slightly so that the total weights of the incoming and outgoing edges are multiplied to the indegrees and outdegrees of each node,
\be
	\label{eq:eq2.10}
	\begin{aligned}
		C &= \sum_x \begin{bmatrix}
	\sqrt{\frac{1}{\alpha_x+1}} & \sqrt{\frac{\alpha_x}{\alpha_x+1}} \\
	\sqrt{\frac{\alpha_x}{\alpha_x+1}} & -\sqrt{\frac{1}{\alpha_x+1}}
	\end{bmatrix} \otimes \ket{x}\bra{x} \, \text{, where}\\
		\alpha_x & = \frac{\sum_j w_j e_{i_{j}}}{\sum_j w_j e_{i_{j}} + \sum_j w_j e_{o_{j}}} \\
		\beta_x & = \frac{\sum_j w_j e_{o_{j}}}{\sum_j w_j e_{i_{j}} + \sum_j w_j e_{o_{j}}}.
	\end{aligned}
\ee
Here $e_{i_{j}}$ represents the $j^{th}$ incoming edge that has the weight $w_j$. Similarly, $e_{o_{j}}$ is the $j^{th}$ outgoing edge with the corresponding weight $w_j$.

%==================================
\subsection{Summary of Algorithm}
%====================================

For any step at each node, the probability amplitude simultaneously evolves through connecting edges, and hence the nodes with a higher weight of outgoing edges will end up having lower amplitudes on average. A summary of steps involved in our algorithm for ranking nodes in order of implementation is given below:
\begin{enumerate} 
	\item Initial state preparation:\\
 The initial state of the system is set to one of equal superposition, i.e., \\ $\ket{\psi}_0 = \left(\frac{\ket{\uparrow} + \ket{\downarrow}}{\sqrt{2}} \right) \otimes \sum_{x=1}^{N} \frac{1}{\sqrt{N}} \ket{x} $.
	\item Coin operation:\\
 The coin operation [Eq.~\ref{eq:eq2.6}] is performed on the state.
	\item Shift operation:\\
 The shift operation [Eq.~\ref{eq:eq2.7}] is performed on the system.
			\item Repetition and measurement:\\
			One step is defined as a single application of the coin operator followed by the shift operator. The system is initially allowed to run for 50 steps independent of network size so that the convergence of ranks hierarchy is seen. See appendix for our analysis on convergence of the quantum ranks.  It is then probed for probability values after each step, reset, and run again for a higher number of steps. The normalized average of the probability values of each node gives its relative importance, considered to be the 'quantum rank' of that node. 
		
	\end{enumerate}

%==========================================

\section{Ranking nodes on various networks}
\label{sec:res}
%===========================================

In this section, we present the results of testing our algorithm on various networks. We have compared the quantum ranks using D-DTQW with the classical ranks obtained by applying the Google PageRank algorithm\,\cite{BP98, BMPW98a, BMPW98b, LM04, HK03, AM12} on different networks. 
\begin{figure}[!h]
  \includegraphics[width=\linewidth]{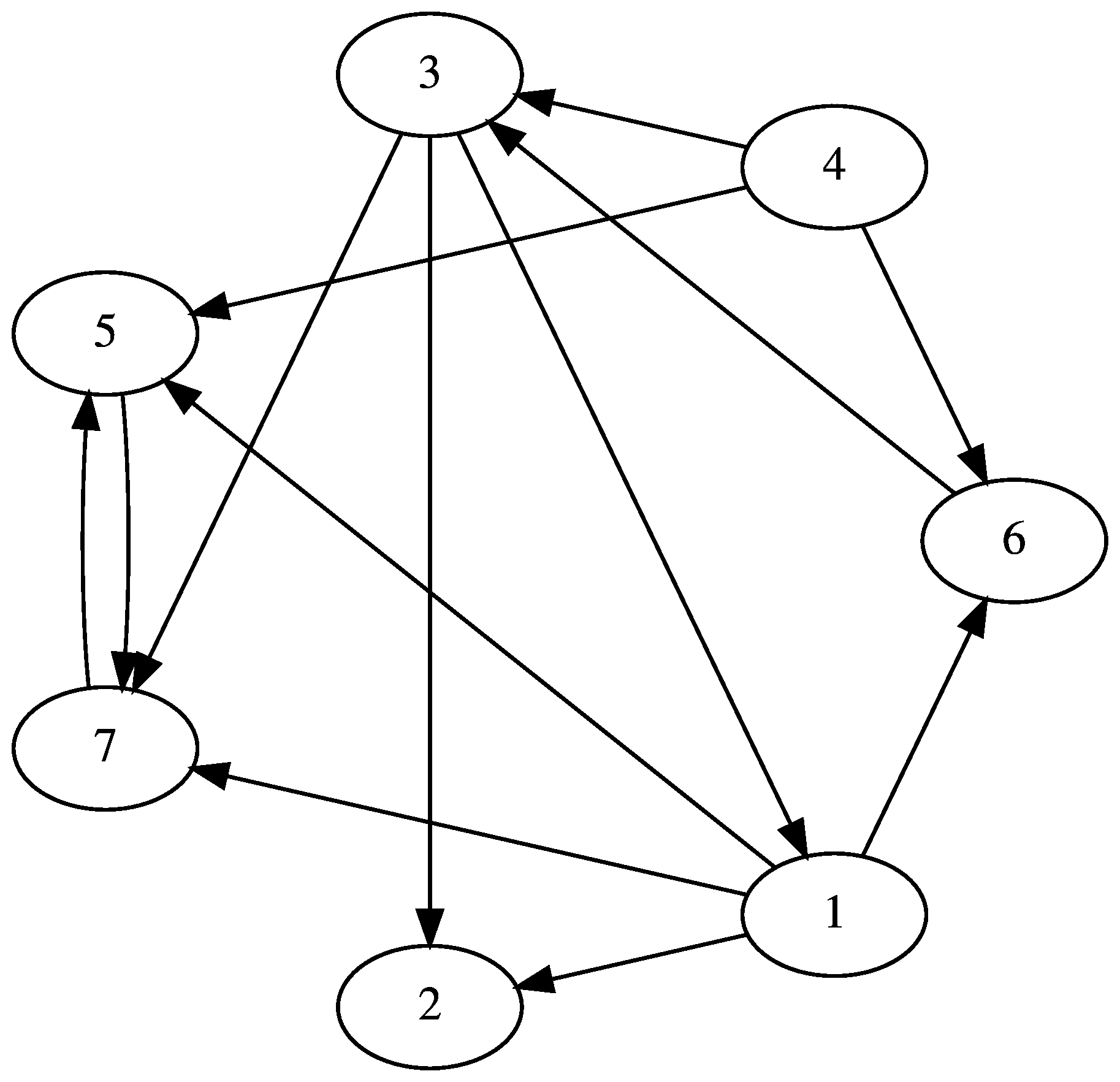}
  \caption{The seven node random network used for testing the algorithm. This network is identical to the one used for presenting quantum page rank in Ref.\,\cite{PMD12}.}
  \label{fig:Graph7Nodes}
\end{figure}

The classical PageRank is calculated by a power method. The initial state is defined as the vector $V = \frac{1}{N} \begin{bmatrix} 1\\ 1 \\ \vdots \\ 1 \end{bmatrix}$. The Google matrix is defined as the column-stochastic matrix $G = (1-p)A + \frac{p}{N}B$, where $A$ is the adjacency matrix of the digraph, $B$ is a square matrix of size $N$, such that all its elements are $1$, and $p$ is a real number lying in $[0,1]$. In this case, we have chosen $p=0.85$. The classical algorithm computes $V^* = G^k V$ with $k=1,2,3,...$, and stops when the value of $V^*$ is the same for $k=k^*$ and $k=k^*+1$. At this point, $V^*$ is an eigenvector of $G$, and the elements of $V^*$ give the classical ranks for the nodes of the network. 

\begin{figure}[!h]
  \includegraphics[width=\linewidth]{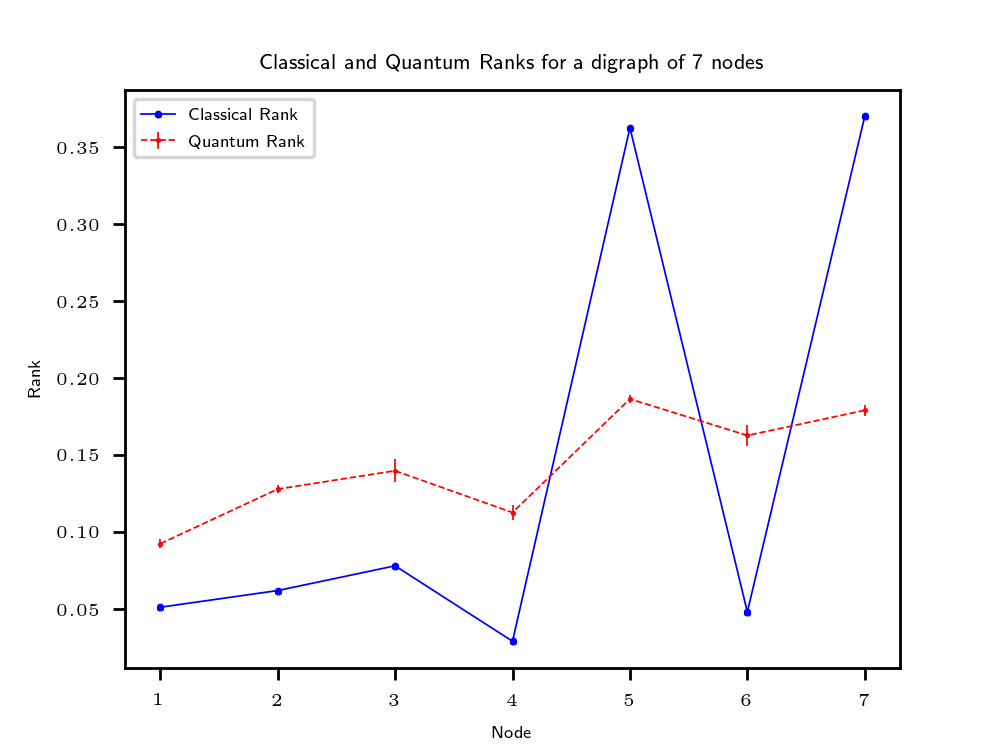}
  \caption{Quantum Ranks of a random 7 Node network shown in Fig.\,\ref{fig:Graph7Nodes} as calculated by the D-DTQW algorithm after 500 steps. As can be seen, the algorithm can identify the various levels of the ranks. The order of the nodes within the levels may, however, be different, as visible here. For the case of this graph, the node with least classical importance is also the node with the least quantum rank. }
  \label{fig:Gen7Nodes}
\end{figure}

In Fig.\,\ref{fig:Graph7Nodes} we have shown a cyclic digraph network with seven nodes. This network is identical to the one used for presenting quantum PageRank algorithm based on quantization of Markov chains in Ref.\,\cite{PMD12}. In Table\,\ref{tab:table1} the results obtained using of using classical ranking algorithm and D-DTQW algorithm after 500 steps is shown. Plot with the values at each node for the network are shown in Fig.\,\ref{fig:Gen7Nodes}. Though the corresponding values at each node is different for classical and quantum algorithms, the returned ranks are identical. 
 \begin{centering}
\begin{table}[!h]
\begin{comment}
	\begin{tabular}{c|c|c|c}
	    \textbf{Node} & \textbf{Classical Rank} & \textbf{DTQW  Rank (variance)} \\
	    \hline
	     & & &\\
	    1 & 0.05108698 & 0.0890760 (0.0021759) &0.0894895 (0.0025600)\\
	    2 & 0.06194378 & 0.1265460 (0.0050376) &0.1267510 (0.0025688)\\
	    3 & 0.07804179 & 0.1305870 (0.0040337) &0.1401660 (0.0078290)\\
	    4 & 0.02898617 & 0.0765860 (0.0014675) &0.1085886 (0.0050009)\\
	    5 & 0.36304910 & 0.2176910 (0.0111097) &0.1901050 (0.0019504)\\
	    6 & 0.04805163 & 0.1313450 (0.0049477) &0.1624721 (0.0942614)\\
	    7 & 0.37056722 & 0.2281690 (0.0105490) &0.1824250 (0.0036141)\\
	\end{tabular}
\end{comment}
\begin{tabular}{c|c|c}
	    \textbf{Node} & \textbf{Classical Rank} & \textbf{DTQW  Rank (variance)} \\
	    \hline
	    1 & 0.05108698 & 0.0894895 (0.0025600)\\
	    2 & 0.06194378 & 0.1267510 (0.0025688)\\
	    3 & 0.07804179 & 0.1401660 (0.0078290)\\
	    4 & 0.02898617 & 0.1085886 (0.0050009)\\
	    5 & 0.36304910 & 0.1901050 (0.0019504)\\
	    6 & 0.04805163 & 0.1624721 (0.0942614)\\
	    7 & 0.37056722 & 0.1824250 (0.0036141)\\
\end{tabular}
\caption{Results of our scheme on the network shown in Fig.\,\ref{fig:Graph7Nodes}}
\label{tab:table1}
\end{table}
\end{centering}

\begin{figure}[!h]
		\includegraphics[width=\linewidth]{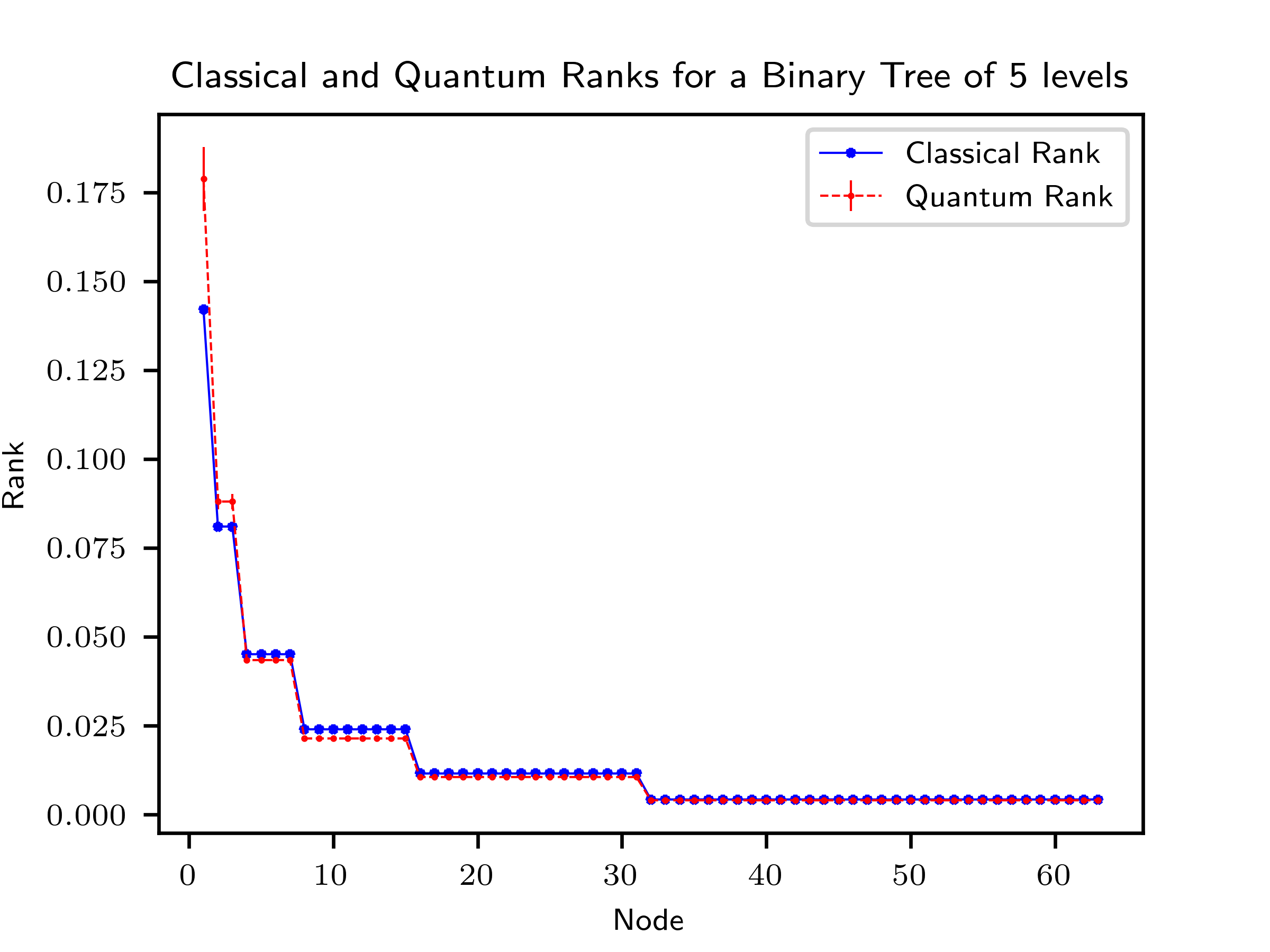}
  		\caption{Quantum rank at each node on a 5-level binary tree network. Quantum rank is consistent with the classical ranking and it efficiently identifies each level of the tree. As expected, the quantum ranks of the nodes in the same level are all equal.}
  \label{fig:Tree63Nodes}
\end{figure}

\begin{figure}[!h]
		\includegraphics[width=\linewidth]{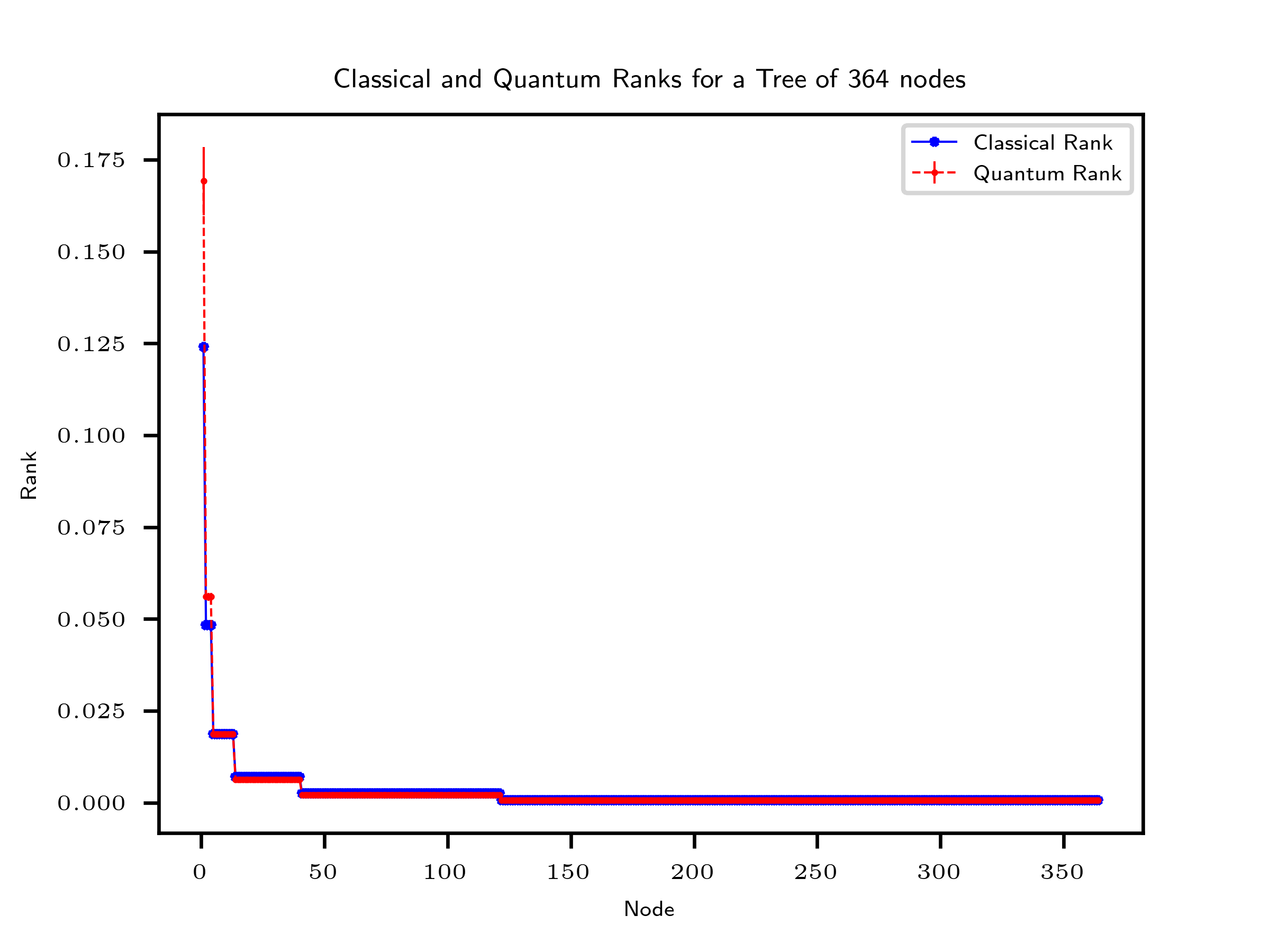}
  		\caption{Quantum ranks for a 5-level tree with branching ratio 3. As with the binary tree, our algorithm is able to successfully discern different levels of this tree. The nodes on the same level have the same ranks, as expected.}
  \label{fig:Tree364Nodes}
\end{figure}

We have simulated our algorithm on a tree network of different levels. An example of five-tree network with 63 nodes is shown in Fig.\,\ref{fig:Tree63Nodes}. This algorithm gives very accurate results on this type of network model, and demonstrates a very nice scalability as it identifies the levels extremely well. The quantum and classical ranks of the different levels are shown in tables~\ref{tab:table2}-\ref{tab:table3} and corresponding plots are shown in Figs.~\ref{fig:Tree63Nodes}, and ~\ref{fig:Tree364Nodes}.
\begin{figure}[!h]
		\includegraphics[width=\linewidth]{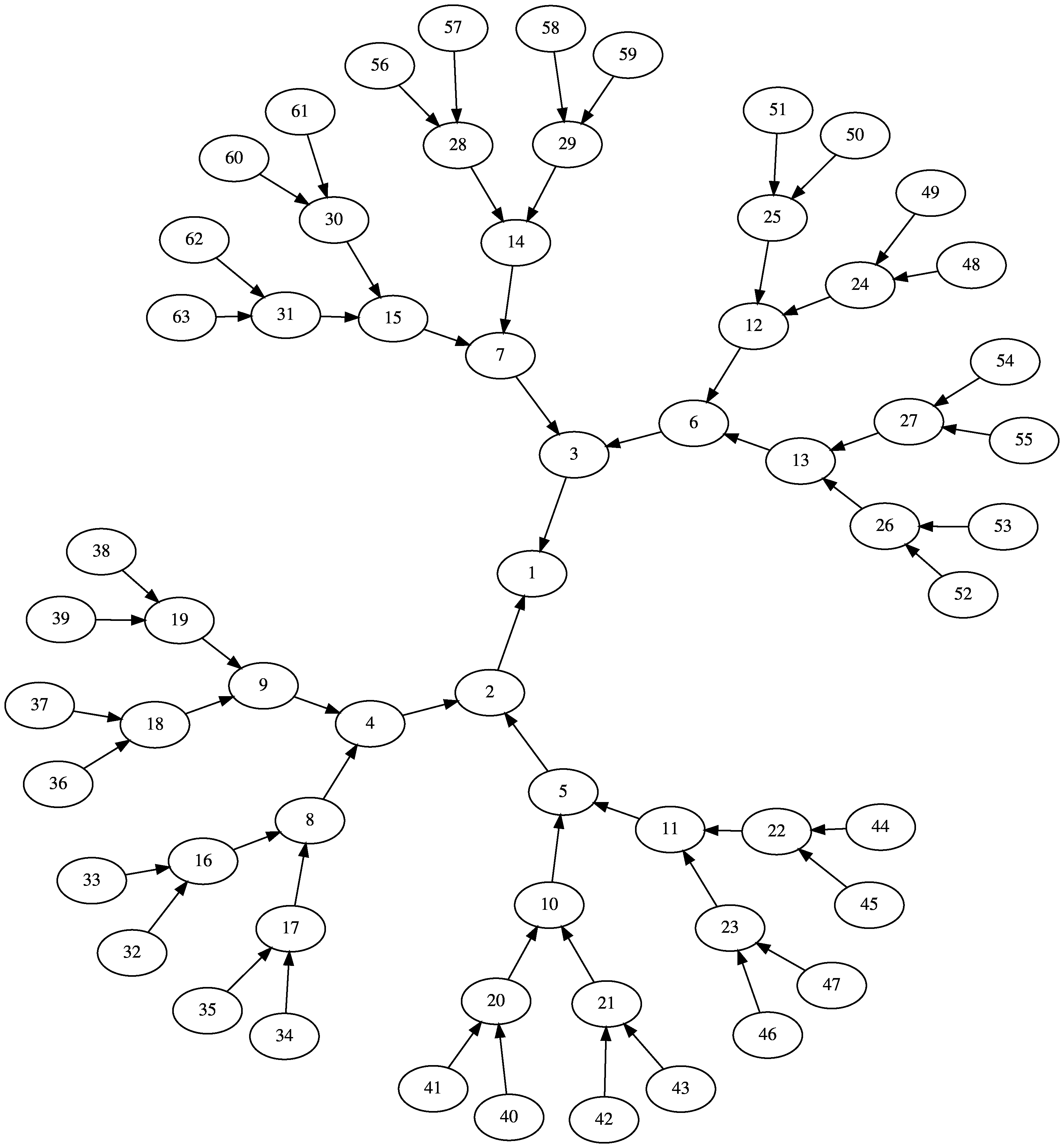}
		\caption{The 5-level binary tree network used for testing our algorithm.}
	\label{fig:TreeGraph63Nodes}
\end{figure}		
\begin{centering}
\begin{table}[!h]
\begin{tabular}{c|c|c}
	    \textbf{Tree Level} & \textbf{Classical Rank} & \textbf{DTQW  Rank (variance)} \\
	    \hline
	    1 & 0.14205971 & 0.1794099 (0.008976)\\
	    2 & 0.08103215 & 0.0878046 (0.002109)\\
	    3 & 0.04513478 & 0.0432301 (0.000561)\\
	    4 & 0.02402133 & 0.0214871 (0.000134)\\
	    5 & 0.01160276 & 0.0106261 (0.000003)\\
\end{tabular}
\caption{Results of our scheme on the network shown in Fig.\,\ref{fig:Tree63Nodes}}
\label{tab:table2}
\end{table}
\end{centering}
		
\begin{centering}
\begin{table}[!h]
\begin{tabular}{c|c|c}
	    \textbf{Tree Level} & \textbf{Classical Rank} & \textbf{DTQW  Rank (variance)} \\
	    \hline
	    1 & 0.12407655 & 0.178774 (0.012538)\\
	    2 & 0.04836995 & 0.057356 (0.000920)\\
	    3 & 0.01868464 & 0.018113 (0.000009)\\
	    4 & 0.00705091 & 0.005964 (0.000001)\\
	    5 & 0.00459601 & 0.004013 (0.000001)\\
\end{tabular}
\caption{Results of our scheme on the network shown in Fig.\,\ref{fig:Tree364Nodes}}
\label{tab:table3}
\end{table}
\end{centering}

\begin{figure}[!h]
		\includegraphics[width=\linewidth, angle=90]{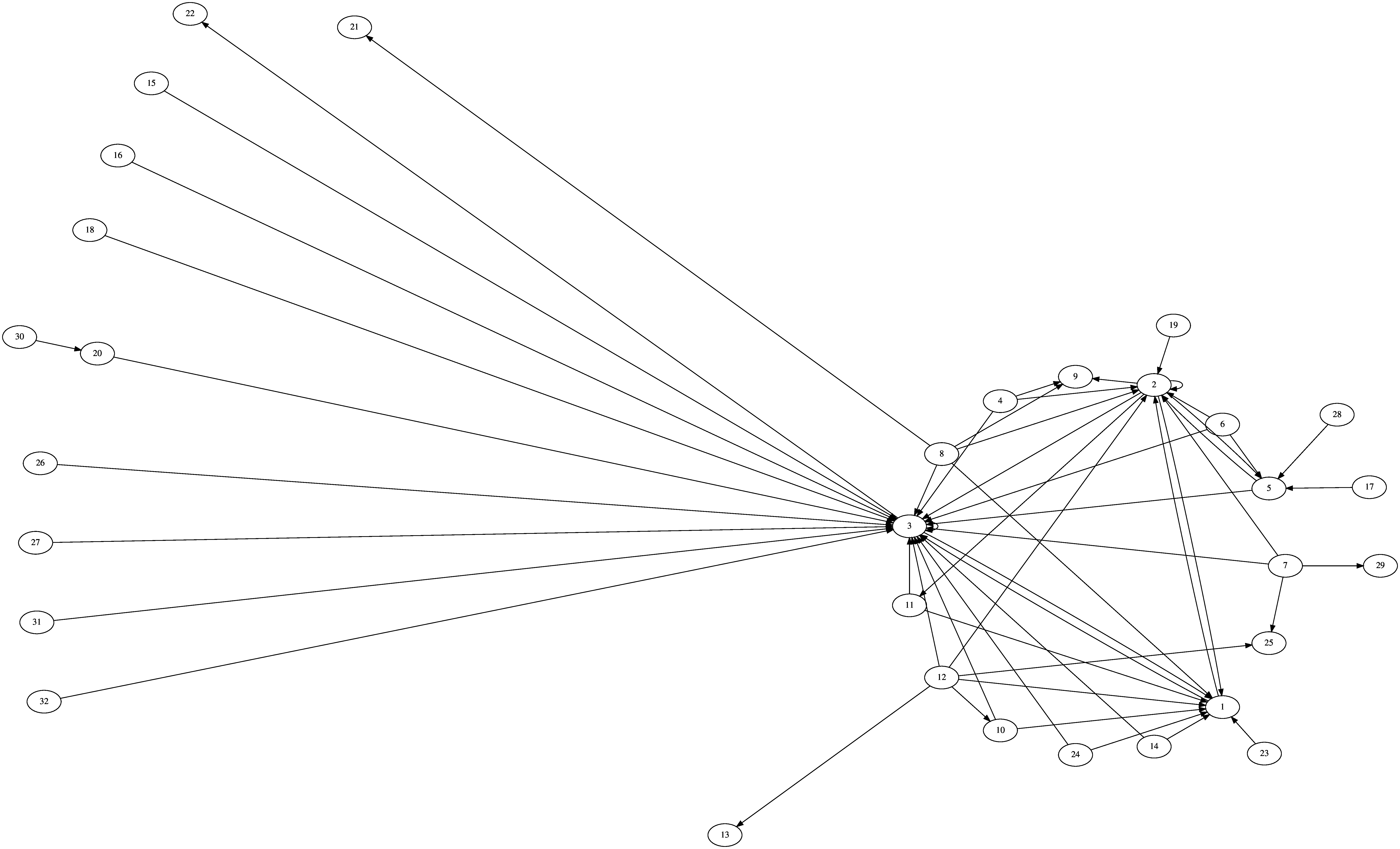}
  		\caption{The 32-node scale-free network used for testing.}
  \label{fig:Graph32NodesPlot}
\end{figure}

\begin{figure}[!h]
		\includegraphics[width=\linewidth]{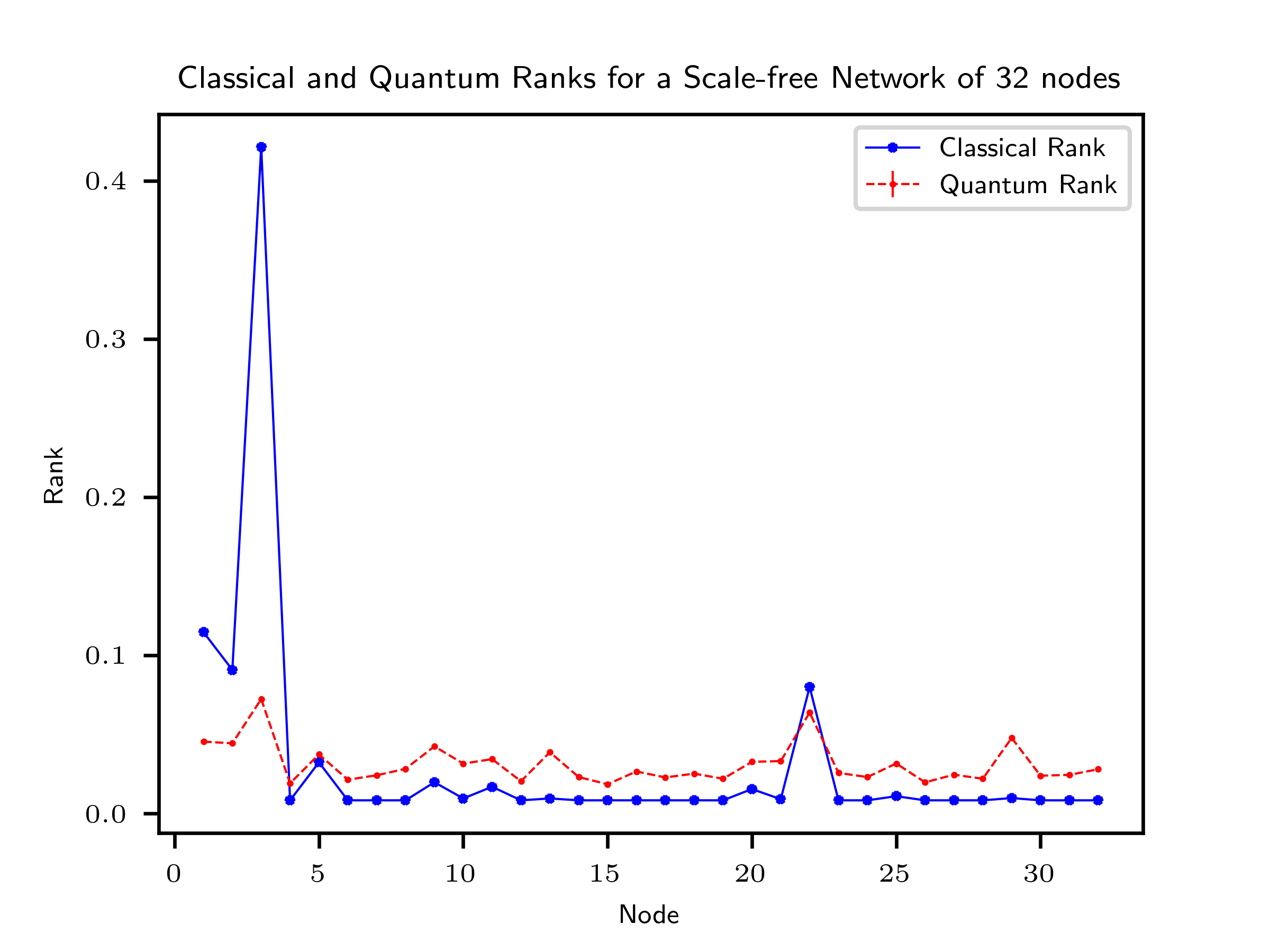}
  		\caption{Quantum ranks obtained by our algorithm on a 32-node scale-free network, plotted against the classical rankings. The network we have tested it on is shown in Fig.\,\ref{fig:Graph32NodesPlot}. While highest ranked node in the quantum protocol is also the highest classically ranked node, the ranks of the intermediate nodes violate the classical orderings.}
  \label{fig:Ranks32Nodes}
\end{figure}

The scheme also gives reasonable output for scale-free networks. Testing was done on a 32-node network shown in Fig.\,\ref{fig:Graph32NodesPlot}, and the results are shown in Fig.\,\ref{fig:Ranks32Nodes}. Even though the quantum ranks do not match the classical ones as exactly as in the case of directed tree networks, they still are able to correctly find the most important node, which in this case, is node 3.

The algorithm is also able to successfully identify the classical nodes with relatively smaller importances, namely nodes 1, 2, 22, 5, 9 and 11. The quantum ranks of these nodes are different from their classical counterparts, and the classically expected hierarchy is also violated in this case due to quantum fluctuations.

\begin{centering}
\begin{table}[!h]
\begin{tabular}{c|c|c}
	    \textbf{Node} & \textbf{Classical Rank} & \textbf{DTQW  Rank (variance)} \\
	    \hline
	    3  & 0.42131700 & 0.07093110 (0.00166186) \\
	    1  & 0.11467604 & 0.04520861 (0.00100160)\\
	    2  & 0.09082545 & 0.04662555 (0.00107079)\\
	    22 & 0.08001230 & 0.06313678 (0.00185716)\\
	    5  & 0.03241485 & 0.03785673 (0.00092886)\\
        9  & 0.01981836 & 0.04244159 (0.00104070)\\
        11 & 0.01696630 & 0.03394764 (0.00059992)\\
\end{tabular}
\caption{Results of our scheme on the network shown in Fig.\,\ref{fig:Graph32NodesPlot}}
\label{tab:table4}
\end{table}
\end{centering}


\begin{figure}[!h]
		\includegraphics[width=\linewidth]{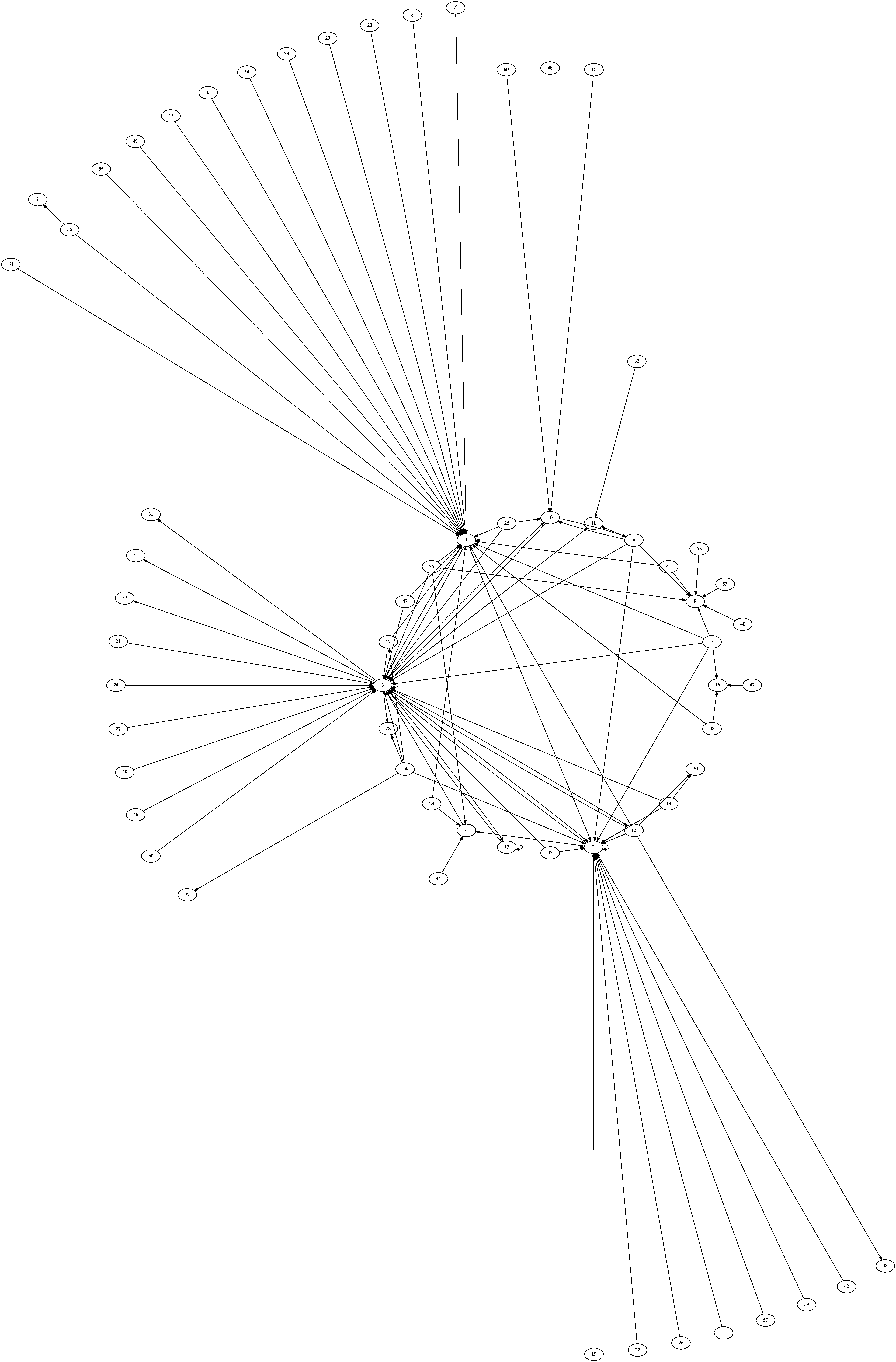}
  		\caption{The 64-node scale-free network used for testing.}
  \label{fig:Graph64NodesPlot}
\end{figure}

\begin{figure}[!h]
		\includegraphics[width=\linewidth]{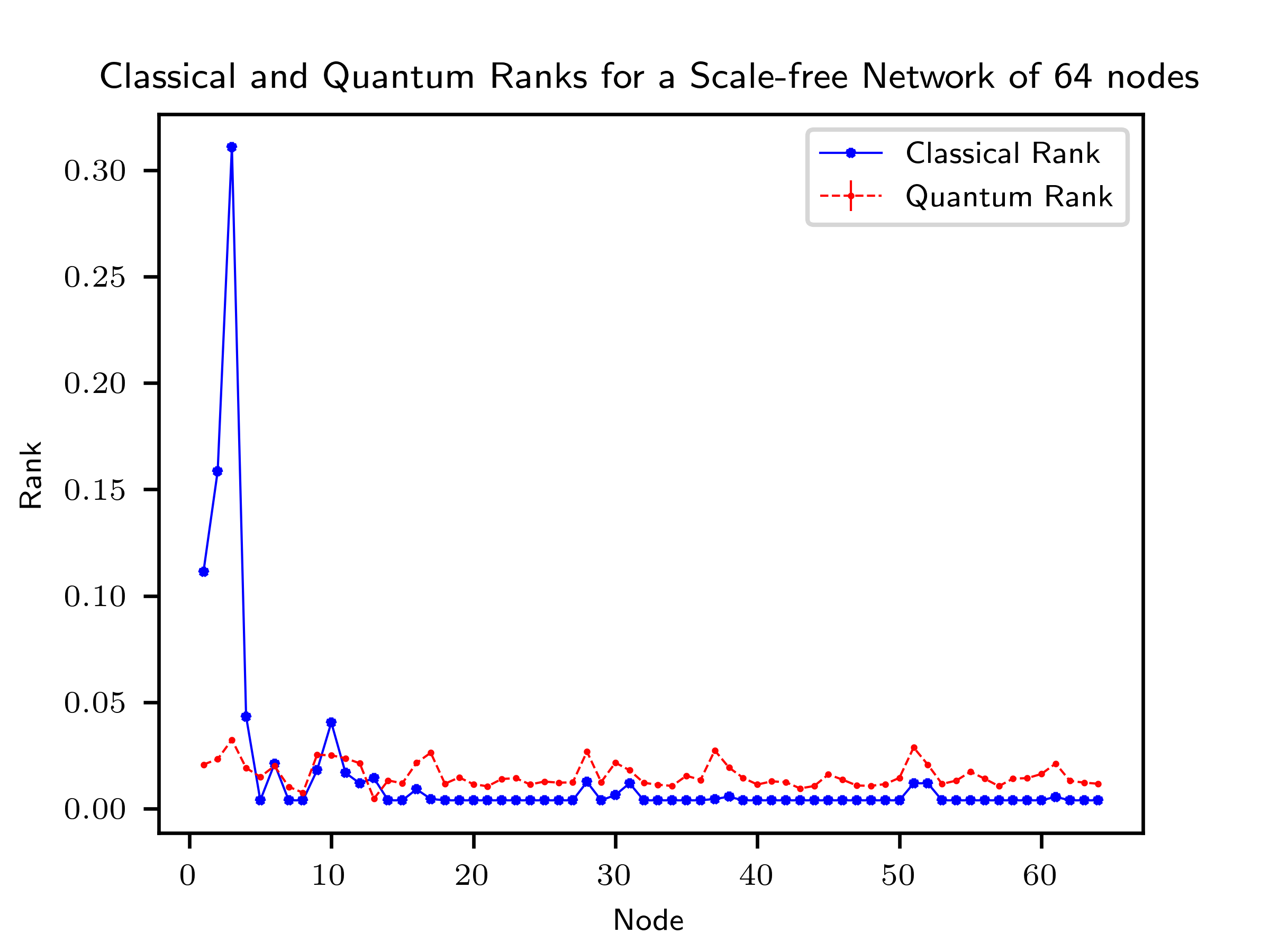}
  		\caption{Quantum ranks of a 64-node scale-free network. The network is shown in Fig.\,\ref{fig:Graph64NodesPlot}. As with the 32-node case, the classically highest ranked node is also the highest ranked node in the quantum protocol, but the milder fluctuations in classical ranks are enhanced in the quantum case, and the intermediate-ranked nodes again violate the classically determined hierarchy.}
  \label{fig:Ranks64Nodes}
\end{figure}

To demonstrate scalability on a scale-free network, we have tested our algorithm on a 64-node scale-free network as shown in Fig.\,\ref{fig:Graph64NodesPlot}, and again as for the 32-node network case, it violated the classical hierarchy to some extend as Fig.\,\ref{fig:Ranks64Nodes}, but it identified the most important node to be the same as the classical case, i.e., node 3. It may be observed that at places where the classical ranks show only a very minuscule change, the quantum ranks have a more exaggerated result, highlighting the importance of the node.
\begin{centering}
\begin{table}[!h]
\begin{tabular}{c|c|c}
	    \textbf{Node} & \textbf{Classical Rank} & \textbf{DTQW  Rank (variance)} \\
	    \hline
	    3  & 0.31100452 & 0.03272801 (0.00036173)\\
	    2  & 0.15862896 & 0.02166889 (0.00021582)\\
	    1  & 0.11133144 & 0.02141180 (0.00022876) \\
	    4  & 0.04330243 & 0.01518750 (0.00010749) \\
	    10 & 0.04044220 & 0.02338800 (0.00027185) \\
        6  & 0.02104263 & 0.01867057 (0.00018800) \\
        9  & 0.01804840 & 0.02568506 (0.00038090) \\
\end{tabular}
\caption{Results of our scheme on the network shown in fig.\ref{fig:Graph64NodesPlot}}
\label{tab:table5}
\end{table}
\end{centering}

\begin{figure}[!h]
		\includegraphics[width=\linewidth, angle=90]{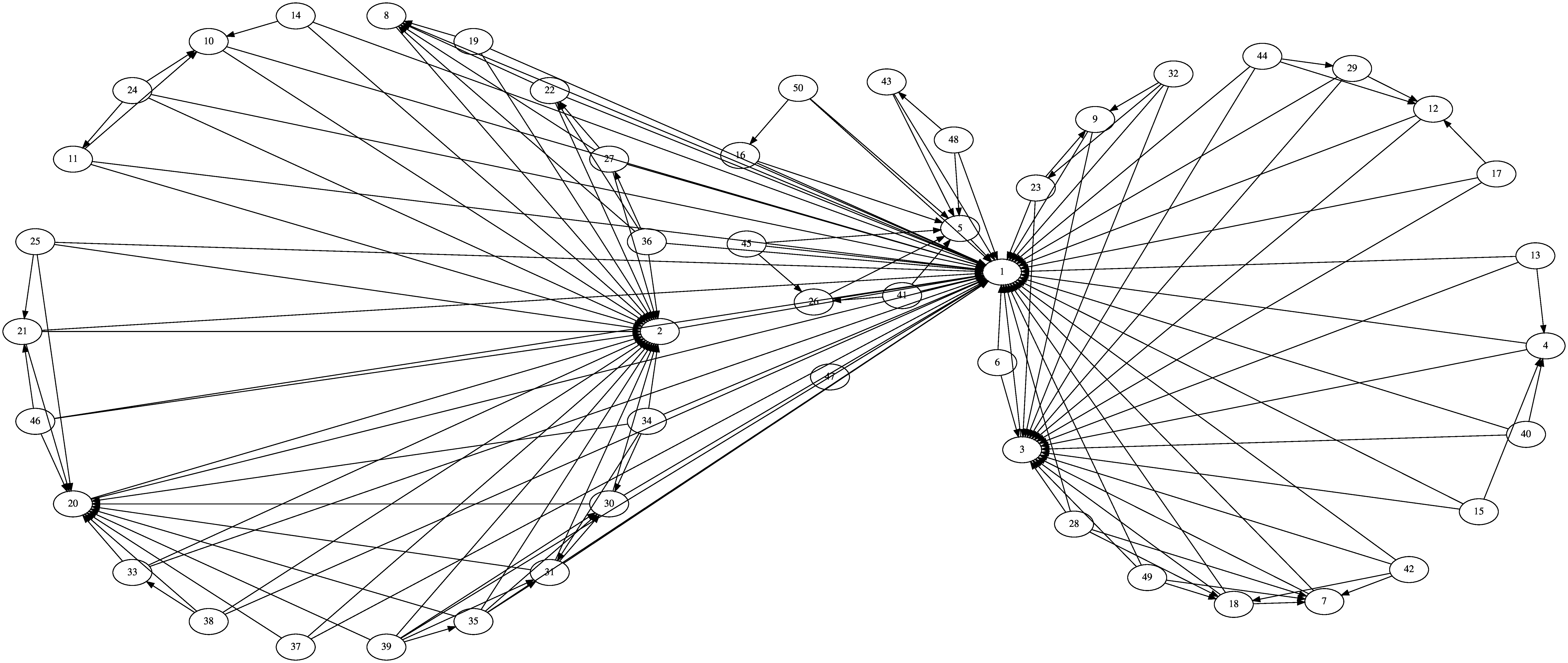}
  		\caption{The 50-node GNC network used for testing.}
  \label{fig:GNCgraphPlot}
\end{figure}
\begin{figure}[!h]
		\includegraphics[width=\linewidth]{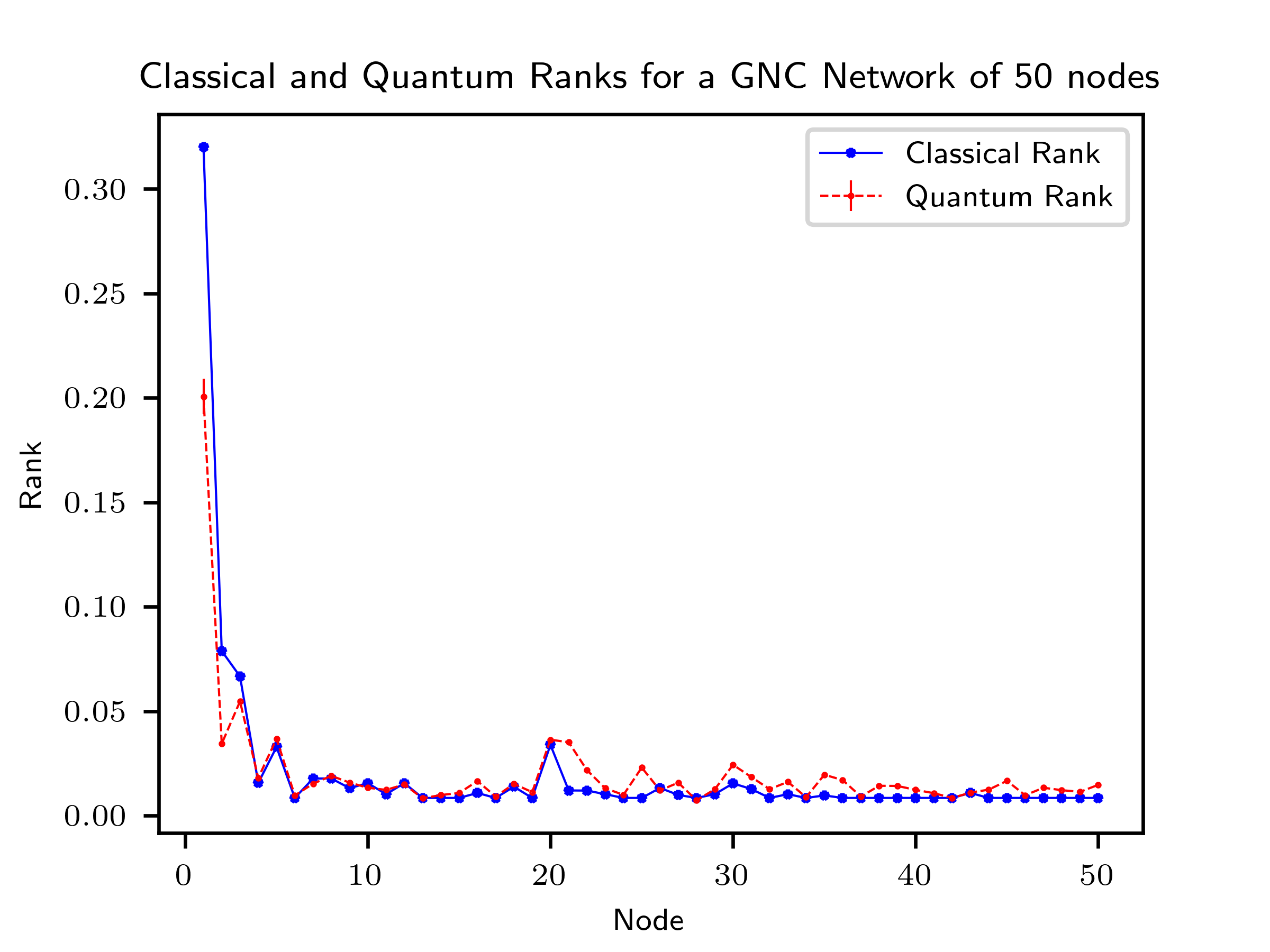}
  		\caption{Plot of the quantum and classical ranks of nodes in a 50-node GNC network, shown in Fig.\,\ref{fig:GNCgraphPlot}. The quantum ranks in this case follow the classical ranks fairly closely, and identify the highest ranked node very quickly. The quantum ranks of nodes of nearly similar importances show a violation of the classically determined order.}
  \label{fig:GNCgraphRanks}
\end{figure}

The algorithm also performs well on other types of digraphs, such as growing networks with copying\,\cite{KR05}. Fig.\,\ref{fig:GNCgraphPlot} presents one such example. The results are shown in Fig.\,\ref{fig:GNCgraphRanks}. Again, the most important nodes are the same as identified in the classical case, but the hierarchy is violated for the other nodes of intermediate and low importances. The classical and quantum ranks for the top-ranked nodes are shown in table~\ref{tab:table4}. It can be seen that the order in this case was not violated for very important nodes.

\begin{centering}
\begin{table}[!h]
\begin{tabular}{c|c|c}
	    \textbf{Node} & \textbf{Classical Rank} & \textbf{DTQW  Rank (variance)} \\
	    \hline
	    1 & 0.32015058 & 0.2012710 (0.0088581)\\
	    2 & 0.07881006 & 0.0343706 (0.0005040)\\
	    3 & 0.06664595 & 0.0547777 (0.0013040)\\
	    20& 0.03386290 & 0.0372421 (0.0006670)\\
	    5 & 0.03284027 & 0.0361666 (0.0005380)\\
\end{tabular}

\caption{Some of the ranks obtained from applying our scheme on the network shown in Fig.\ref{fig:GNCgraphPlot}.}
\label{tab:table6}
\end{table}
\end{centering}

In our results, for all the networks we have chosen, we have run this algorithm for 500 iterations. However, we have found that the order of the nodes does not change after about 200 steps. Any number of steps done post this point only serves to reduce the standard deviation of the quantum rank of the node (See Appendix).

%====================================================================
\section{Conclusions}
\label{sec:conc}
%====================================================================

In this work, we have presented a new algorithm to generate a ranking of the nodes of a network by using a D-DTQW on a network. We use a scattering form of the shift operator, which is derived from the Google matrix used for the PageRank algorithm \,\cite{GGS10, CV07, BP98, BMPW98a, BMPW98b, LM04, HK03, AM12}. The google matrix is derived from the adjacency matrix of the graph, and therefore contains information inherent to the structure of the network within itself. We have shown the results obtained after iterating our algorithm on different networks, and the conclusions that we derive from them are as follows: 
\begin{enumerate}
	\item Directed tree networks :\\
 The D-DTQW algorithm is instantaneously able to figure out the root node, however, the instantaneous values of the quantum ranks for the nodes on larger network violate the expected classical hierarchy (from Google's PageRank algorithm). This is expected as the violation arises as a consequence of quantum fluctuations in the system. Also, since quantum walks do not have steady state solutions, the instantaneous ranks will never converge. The mean values of the quantum ranks, however, obey the classically expected hierarchy. The hierarchy within a particular level is not violated for this case. However, for networks used for quantum communication and quantum processing the hierarchy of nodes returned by the D-DTQW algorithm will be more relevant over the classical hierarchy. The D-DTQW dynamics takes into account all the quantum interference and fluctuations from the quantum dynamics in network.
 
 \item Other random networks:\\
	 Our quantum algorithm also works well for scale-free networks, GNC networks and other random networks (results are shown only for the scale-free and GNC digraphs in this work) identifies the well-connected (i.e., most important) nodes very quickly. The internal order of the hubs and the other nodes on the network may differ from what is classically expected, but the algorithm can separate the two types of nodes. As with the case of the trees, the instantaneous quantum ranks do not converge, however, the averaged values of these instantaneous ranks do. There are nodes, however, for which the averaged quantum ranks also violate the classically expected hierarchy. This indicates the deviation due to quantum interference and fluctuation in the networks.
\end{enumerate}
\par 
From the results listed above, it becomes clear that our algorithm shows some nontrivial features that are also found in the classical PageRank algorithm. For non-trivial networks some deviations of quantum rank from the classical rank highlights the role of quantum interference and quantum fluctuations. From this we can say that the ranking of nodes for network with quantum and classical processing of information may not be identical and quantum scheme is very much required for analysis of architecture and network for quantum processes.   

The quantum scheme however can be mapped to a dynamics on quantum systems and the computation can be experimentally performed on it. This quantum algorithm can inspire for applications in data sciences and other areas wherever there exists a need to generate a ranking of nodes and analyse the networks.

%============================================

\vskip 0.2in
\noindent
{\bf Acknowledgment:}\\
\\
\noindent
CMC and RM would like to thank Department of Science and Technology, Government of India for the Ramanujan Fellowship grant No.:SB/S2/RJN-192/2014. We also acknowledge the support from Interdisciplinary Cyber Physical Systems (ICPS) programme of the Department of Science and Technology, India, Grant No.:DST/ICPS/QuST/Theme-1/2019/1.

%=================================

\newpage

%\begin{widetext}

%========================================
\section{Appendix}
\label{sec:ap}

We demonstrate here the convergence of the quantum ranks for the networks listed above. It is seen that for the case of the random network, the algorithm takes roughly 200 steps to get the correct order of nodes. For the case of the tree networks (and in general with acyclic networks), the algorithm starts giving the correct ranking from step 1 itself. For all other networks (scale-free and GNC networks), the order converges to the final output order after roughly 50 steps, independent of the size of the network. We have run the algorithm on each graph for 500 iterations to make the errors small.

\begin{widetext}

\begin{table}[!h]
	
\begin{tabular}{cc}
%\begin{figure}[!h]
%	\label{fig:Conv7NodesRandom}
	\includegraphics[width=0.4\linewidth]{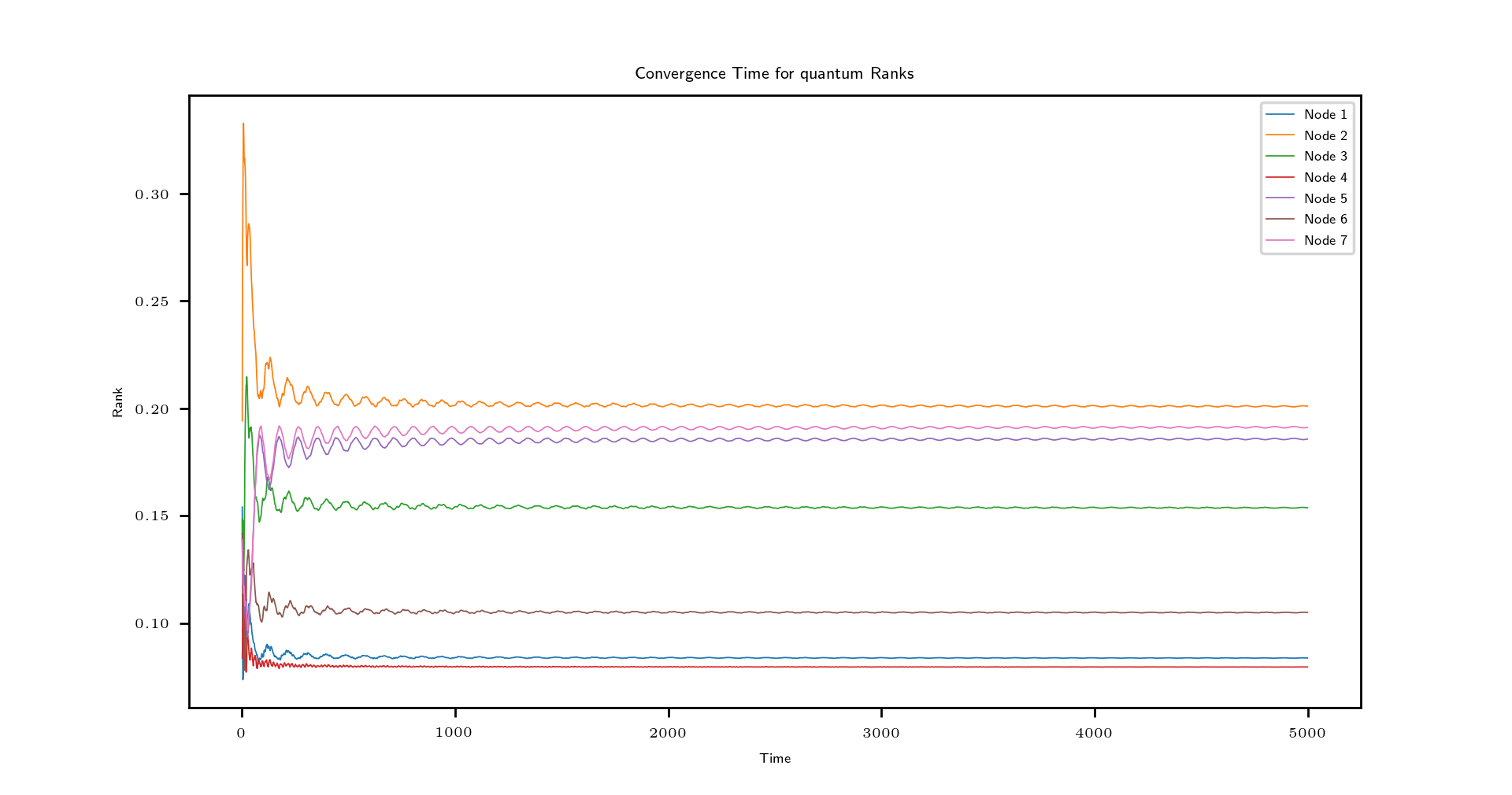}
%	\caption{The convergence of quantum ranks for the 7-node random graph shown in fig.\,\ref{fig:Graph7Nodes}}
%\end{figure} 
&
%\begin{figure}[!h]
%	\label{fig:ConvTree63}
	\includegraphics[width=0.4\linewidth]{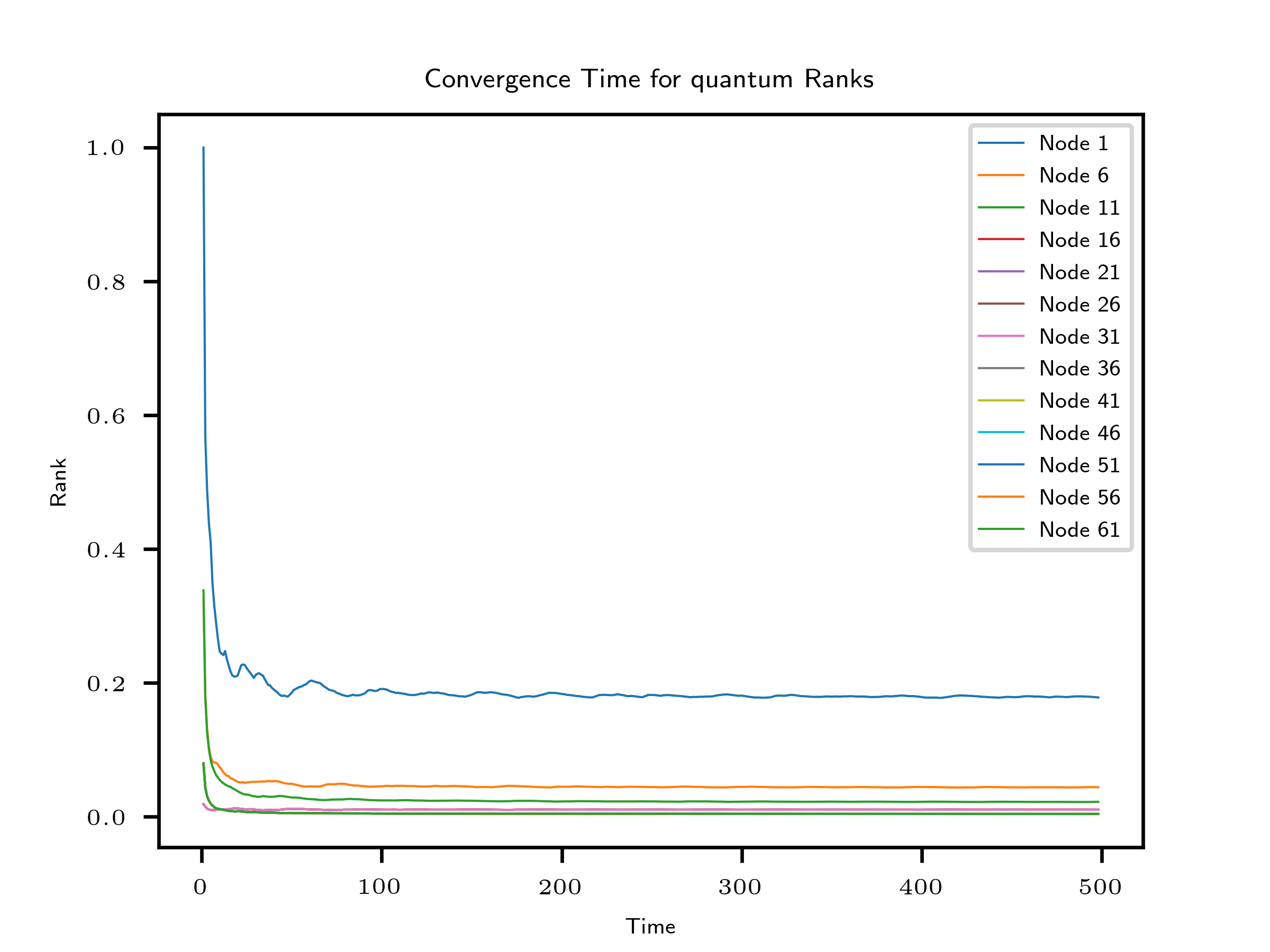}
%	\caption{The convergence of quantum ranks for the 5-level binary tree graph shown in fig.\,\ref{fig:TreeGraph63Nodes}}
%\end{figure} 
\\

(a): Convergences for the network of Fig.\,\ref{fig:Graph7Nodes} & 
(b): Convergences for the network of Fig.\,\ref{fig:TreeGraph63Nodes} \\

%\begin{figure}[!h]
%	\label{fig:ConvTree364}
	\includegraphics[width=0.4\linewidth]{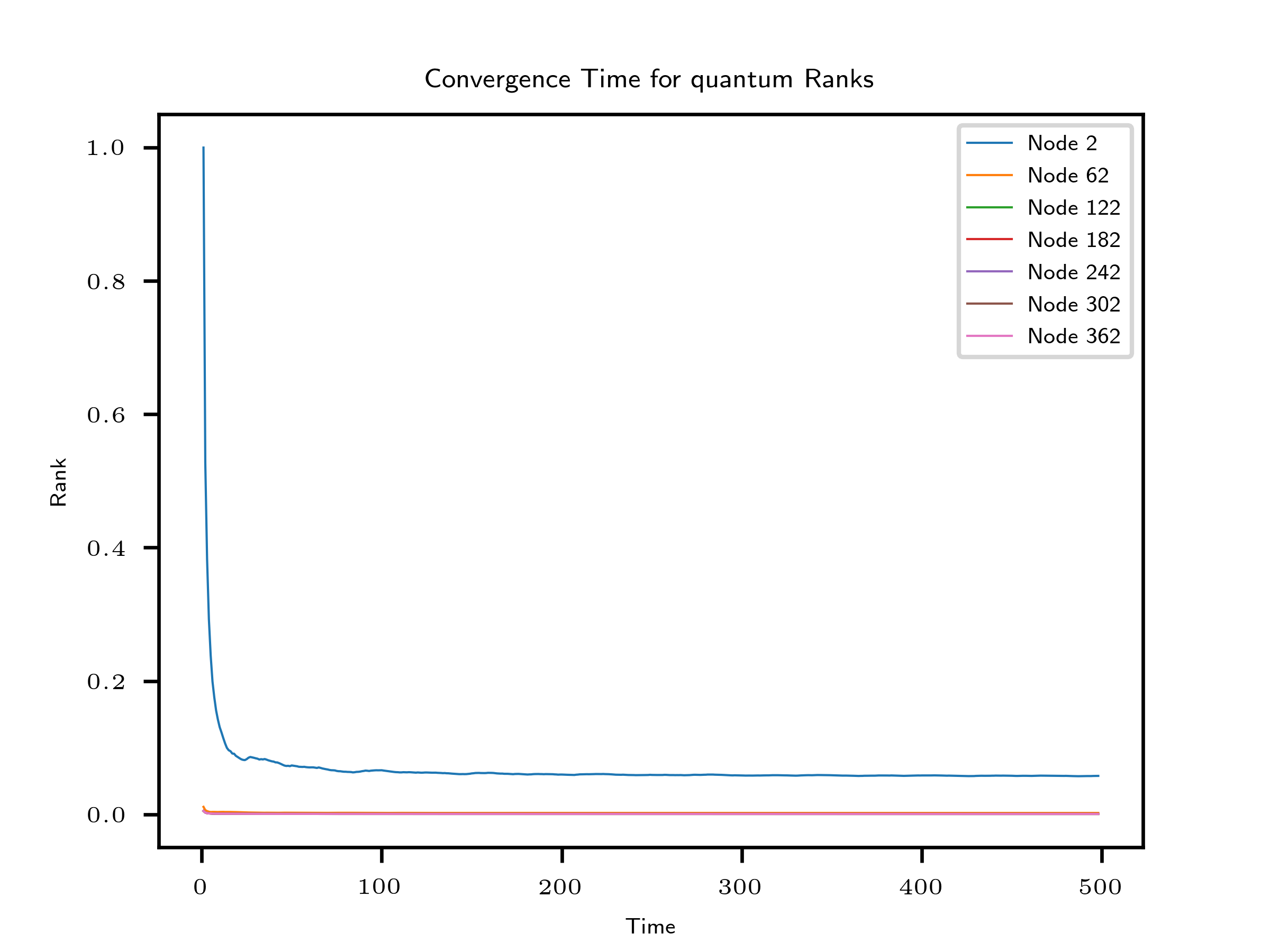}
%	\caption{The convergence of quantum ranks for a 5-level tree graph with branching ratio 3.}
%\end{figure}
 &
%\begin{figure}[!h]
%	\label{fig:ConvScaleFree32}
	\includegraphics[width=0.4\linewidth]{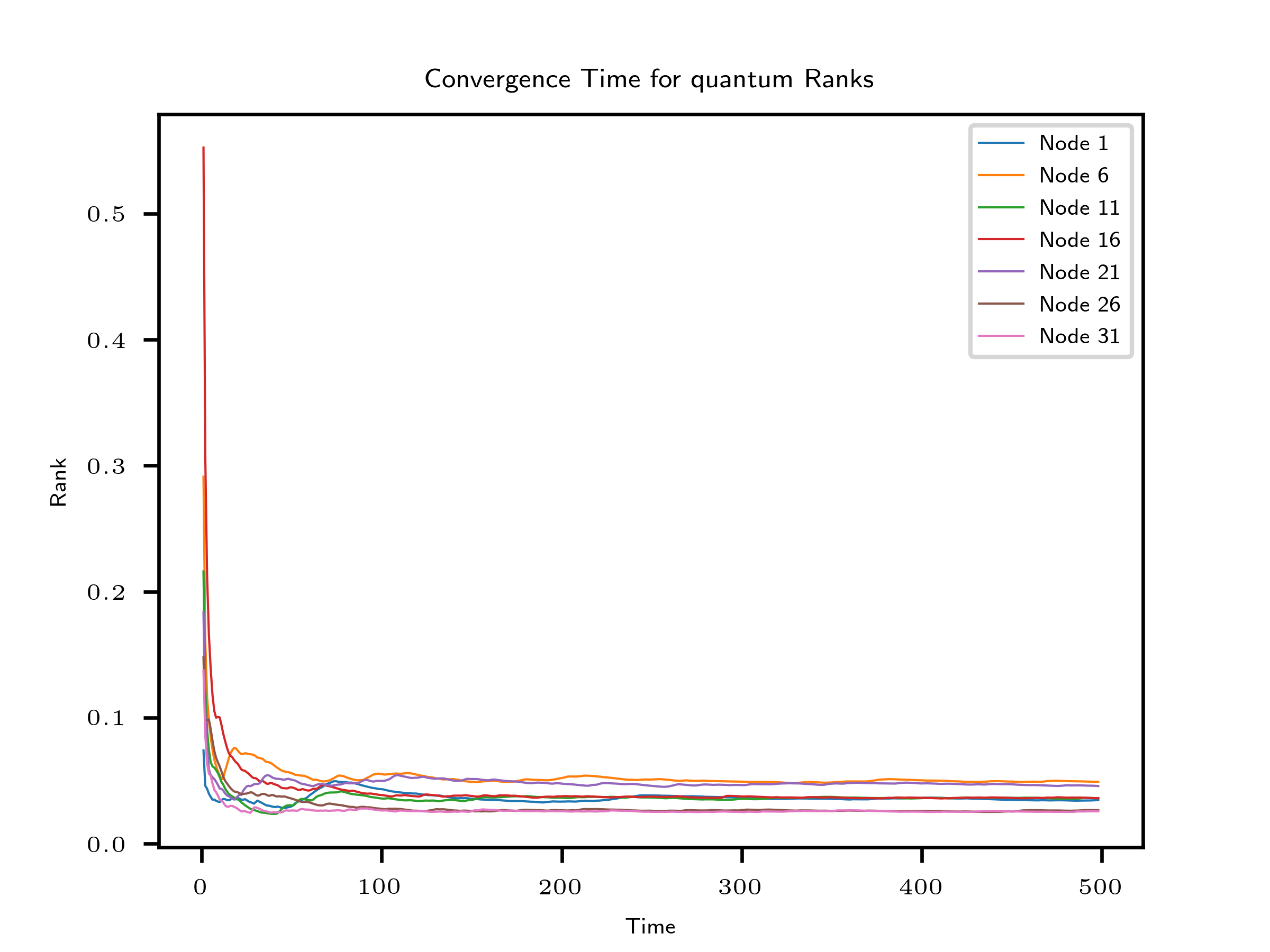}
%	\caption{The convergence of quantum ranks for the 32-node scale-free graph shown in fig.\,\ref{fig:Graph32NodesPlot}}
%\end{figure}
\\
(c): Convergences for the 5-level tree with branching ratio 3. & 
(d): Convergences for the network of Fig.\,\ref{fig:Graph32NodesPlot} \\
%\begin{figure}[!h]
%	\label{fig:ConvScaleFree64}
	\includegraphics[width=0.4\linewidth]{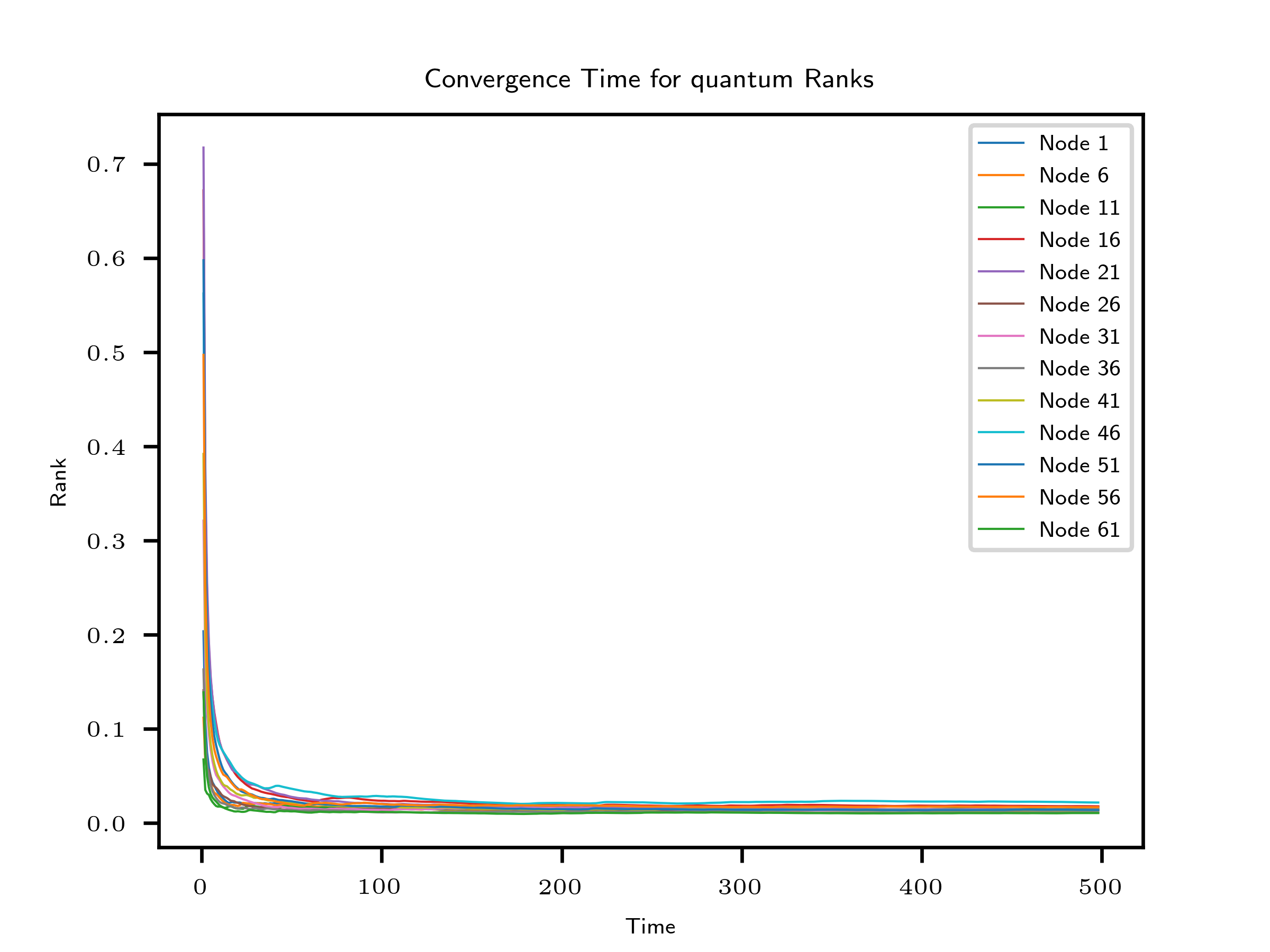}
%	\caption{The convergence of quantum ranks for the 64-node scale-free graph shown in fig.\,\ref{fig:Graph64NodesPlot}}
%\end{figure} 
& 
%\begin{figure}[!h]
%	\label{fig:ConvGNC50}
	\includegraphics[width=0.4\linewidth]{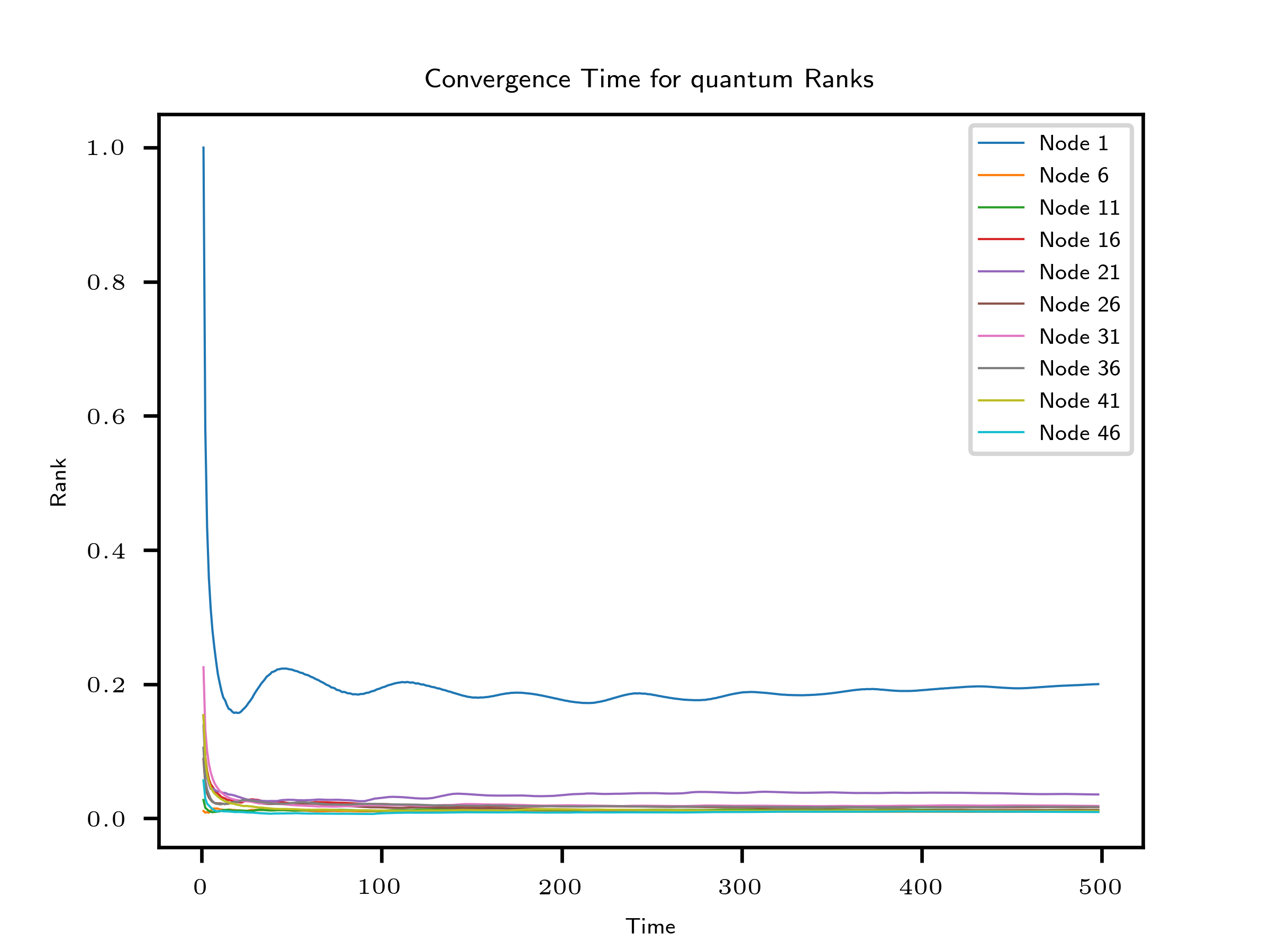}
%	\caption{The convergence of quantum ranks for the 50-node GNC graph shown in fig.\,\ref{fig:GNCgraphPlot}}
%\end{figure} 
\\
(e): Convergences for the network of Fig.\,\ref{fig:Graph64NodesPlot} & 
(f): Convergences for the network of Fig.\,\ref{fig:GNCgraphPlot} \\

\end{tabular}
\end{table}

\centering{Plots of quantum ranks with time to prove convergence. For plots where the curves are not visible (e.g., in plot (e)), they are all coinciding with each other as they have tiny values that are very close to each other.}

\end{widetext}

\end{document}